\documentclass{article}
\usepackage{emulateapj,psfig}

\def\etal{{\it et al.}$\,\,$}

\def\simlt{\mathrel{\spose{\lower 3pt\hbox{$\mathchar"218$}}
     \raise 2.0pt\hbox{$\mathchar"13C$}}}
\def\simgt{\mathrel{\spose{\lower 3pt\hbox{$\mathchar"218$}}
'     \raise 2.0pt\hbox{$\mathchar"13E$}}}
\def\gsim{ \lower .75ex \hbox{$\sim$} \llap{\raise .27ex \hbox{$>$}} }
\def\lsim{ \lower .75ex \hbox{$\sim$} \llap{\raise .27ex \hbox{$<$}} }

\accepted{}

\long\def\***#1{{\scshape ***#1***}}

\makeatletter

\newenvironment{inlinefigure}{%
\def\@captype{figure}%
\noindent\begin{minipage}{0.999\linewidth}\begin{center}}
{\end{center}\end{minipage}\smallskip}
\makeatother

\lefthead{Macci\`o et al.}
\righthead{Constraining $\Lambda$ using Cluster Quadrupoles}

\def\etal{ {et al.}}

\def\be{\begin{equation}}
\def\ee{\end{equation}}
\def\bea{\begin{eqnarray}}
\def\eea{\end{eqnarray}}
\def\mincir{\raise -2.truept\hbox{\rlap{\hbox{$\sim$}}\raise5.truept
\hbox{$<$}\ }}
\def\magcir{\raise -4.truept\hbox{\rlap{\hbox{$\sim$}}\raise5.truept
\hbox{$>$}\ }}
\def\lsim{\raise -2.truept\hbox{\rlap{\hbox{$\sim$}}\raise5.truept
\hbox{$<$}\ }}
\def\gsim{\raise -4.truept\hbox{\rlap{\hbox{$\sim$}}\raise5.truept
\hbox{$>$}\ }}
\def\lcdm{{$\Lambda$CDM} }
\def\Msun{M_\odot}
\def\Msunh{{h^{-1} M_\odot}}
\def\Mpc{\,{\rm Mpc}}
\def\Mpch{\,h^{-1}{\rm Mpc}}

\begin{document}
\title{Constraining $\Lambda$ using Cluster Quadrupoles}

\author{Andrea Macci\`o\altaffilmark{1}, Alessandro Gardini\altaffilmark{2}, 
Sebastiano Ghigna\altaffilmark{1,3} \& Silvio Bonometto\altaffilmark{1,3}}
\altaffiltext{1}{Dipartimento di Fisica G. Occhialini -- Universit\`a 
di Milano Bicocca, Milano, Italy}
\altaffiltext{2}{Dipartimento di Astronomia, Università di Padova, vicolo dell'Osservatorio 5, I-35122 Padova, Italy}
\altaffiltext{3}{INFN sezione di Milano -- Via Celoria 16, I-20133 Milano, Italy}

\authoremail{Andrea.Maccio@axrialto.uni.mi.astro.it, 
Alessandro.Gardini@pd.astro.it}

 
\begin{abstract}
We examine how the statistics of the quadrupoles of (projected) cluster masses
can discriminate between flat cold dark matter (CDM) universes with or without
a cosmological constant term.
Even in the era of high precision cosmology that 
cosmic microwave background experiments should open soon, 
it is important to devise
self consistency tests of cosmogonic theories tuned at the matter radiation
decoupling epoch  using data from the non--linear evolved universe. We 
build cluster catalogs from two large
volume simulations of a ``tilted'' CDM model and a $\Lambda$CDM model
with cosmic density parameter $\Omega_m=0.35$ and cosmological constant
contribution 
$\Omega_\Lambda=0.65$. From the projected mass distribution of the clusters
we work out the quadrupoles $Q$ and examine their dependence on cluster mass
and the cosmological model. We find that TCDM clusters have systematically
larger quadrupoles than their \lcdm counterpart. The effect is mass 
dependent: massive clusters  ($M\gsim 10^{15}\Msunh$) have quadrupoles
differing by more than 30\% in the two models, while for  
$M\lsim 4\times10^{14}\Msunh$ the difference rapidly drops to $\sim 1$\%. 
Performing a K-S test
of the $Q$ distributions, we estimate that using just the 15 most
massive clusters in the simulation volume ($360\Mpch$ a side) we 
can discriminate 
between TCDM and \lcdm at a confidence level better than 
$99.9$\%.
In the volume probed by exhisting observations, 
there are potentially several hundred clusters with masses above
the threshold for which the differences in the quadrupoles become
relevant.
Should weak lensing data become available for this whole set,
a quadrupole analysis may be expected to discriminate among
different values of $\Lambda$.

\vskip 0.2 truecm


\end{abstract}

                                                                                
\section{Introduction}

Clusters of galaxies have been systematically studied in recent 
years, using both optical and X--ray data. In principle, it is possible to 
obtain from their properties stringent  constraints to cosmological models,
e.g. from their mass function 
(e.g. \cite{Eke1,Eke2,Col,Via,Mo,Bor1,Gir,Pos,Gar}).

Optical and X-ray studies have shown that many clusters have complex 
morphologies with strong evidence of substructure (e.g. Geller \& Beers 1982;
Dressler \& Schectman 1988; West \& Bothun 1990; Forman \& Jones 1990; 
Bird 1994; West, Jones \& Forman 1995; Bardelli \etal 1998; Solanes \etal 
1999). 
In parallel to the development of these observations of the visible (barionic)
matter in clusters, the study of the distortions of the images of distant
galaxies gravitationally lensed by cluster potentials 
have made it possible to systematically
map the distribution of the  mass, including
the dynamically dominant dark matter component
(e.g. Tyson, Valdes \& Wrenk 1990; Kaiser \& Squires 1993; Seitz \& Schneider 
1995; for ongoing surveys, e.g. Clowe \etal 2000, 
Dahle 2000, Graham \etal  2000, 
Wittman \etal 2000; for 
recent reviews Hattori, Kneib \& Makino 1999, and Kaiser 1999).

In hierarchical bottom up cosmological models, like the cold dark matter model
(CDM) and its variants, galaxy clusters are the largest bound  
structures to form (White \& Rees 1978; Davis \etal 1985). 
Their properties
can retain the signature
 of the cosmological parameters more easily than smaller older structures.

Indeed, in different cosmologies clusters can have significantly different
formation histories and this may affect their typical morphologies.
This fact was first 
examinated using analytical methods by 
Richstone, Loeb \& Turner (1992), Bartelmann, Ehlers \& Schneider (1993) and
Lacey \& Cole (1993), and, using numerical methods, by Evrard \etal (1994)
Mohr \etal (1995) and Wilson, Cole \& Frenk (1996b; hereafter WCF96).
The latter authors studied the
dependence on cosmology of the quadrupole of a cluster's projected mass by
simulating
 weak gravitational lensing in artificial clusters (Wilson, Cole \& Frenk 
1996a) grown in numerical
simulations of the CDM model with different values of today's
matter density parameter $\Omega_m$ and cosmological constant $\Lambda$.
The relation between the projected mass distribution of clusters and
the cosmological model has also been recently explored in the context of
hydrodynamical simulations by Valdarnini, Ghizzardi \& Bonometto (1999).

In a low-density universe without cosmological constant, the linear growth of
density fluctuations ceases
after a redshift $z\sim (1/\Omega_m)-1$ (e.g. Peebles 1980, secs 11 and 13).
If the model is spatially flat thanks to a non--vanishing cosmological 
constant, the growth of fluctuations freezes
when the vacuum energy density begins to dominate the matter density $\rho_m$,
that is, as $\rho_m\propto (1+z)^3$, at a redshift 
$z\sim (\Omega_\Lambda/\Omega_m)^{1/3} -1$.
Hence, in low-density universes, clusters on average form at moderately high
redshifts ($z\gsim 0.5$) and subsequently accrete little material;
on the contrary, in the standard $\Omega_m=1$ CDM universe, structure formation
occurs continuously with rich galaxy clusters also 
forming in very recent epochs ($z\lsim 0.3$). 
WFC96's analysis showed that the statistics of the
quadrupoles of the
projected mass distribution of clusters is sensitive to the value of
$\Omega_m$, quite independently of the value of $\Lambda$.
 However, other than $\Omega_m=1$, 
they considered only a rather extreme value $\Omega_m=0.2$.
 Also, they examined only a
small sample of (eight) simulated clusters rather than building
a catalog from a large volume simulation and could not examine the
importance of the effect as a function of cluster mass.

In this paper we extend the
study of 
how the statistics of cluster mass quadrupoles is sensitive to the 
cosmological parameters 
building cluster catalogs from
large scale $N$-body simulations. In agreement with recent
data on 
the angular power spectrum of CMB anisotropy (BOOMERANG-98, de Bernardis 
\etal 2000, Lange \etal 2000; MAXIMA-1, Hanany \etal 2000, Balbi \etal 2000; 
see also  White, Scott \& Pierpaoli 2000, Jaffe \etal 2000),
 we consider two flat universes
($\Omega_m + \Omega_\Lambda = 1$). One is a standard cold dark matter 
model with a ``tilt'' in the primordial spectral index ($n=0.8$); this
simulation is the same analyzed by \cite{Gar}.  
The other model has
$\Omega_m=0.35$ and cosmological
constant $\Omega_\Lambda=0.65$ ($\Lambda$CDM), which fits CMB quadrupole and
cluster abundance data and is
compatible with 
the data on high-redshift Type Ia supernovae  which suggest
an accelerating cosmic expansion (\cite{Per,Rie,Teg}).
The TCDM model is ruled out by combined CMB and LSS data 
(e.g. \cite{Teg}),
 but roughly reproduces the 
CMB quadrupole and the observed abundance of rich 
galaxy clusters (e.g. \cite{Gar}). 
This is sufficient for the purposes of this 
analysis, since we want to test $\Lambda$CDM  against an 
example of a standard ($\Omega=1$) CDM cosmology. 

In the next few years, starting from the results of MAP (e.g. Page 2000), 
and also 
thanks to the Planck experiment (e.g. De Zotti \etal 2000), a detailed
reconstruction of the angular spectrum of CMB will be performed.
 Several authors (see, e.g., \cite{Jun,Kam}) have shown that
$\Omega_m$ and $\Lambda$, as well as the spectral
index $n$ of primeval density fluctuation, will then be safely
determined with unprecedented precision. However, not all 
degeneracy
in the parameter space will be removed, unless suitable
{\it a priori} conditions are set. For instance, small components
of hot dark matter, consistent with current limits on
neutrino masses coming from neutrino mixing experiments,
could introduce residual uncertainties. A problem
could also arise if the vacuum energy, accounted for by the cosmological
constant $\Lambda$, is not really constant in time. 
Indeed, interpreting the data on high-redshift supernovae in terms of 
a cosmological {\it constant} may be simplistic. Such data seem to 
require that the energy density is dominated by a negative-pressure
component ({\it dark energy}); but a constant vacuum energy is just one 
possibility. A popular alternative
scenario is quintessence (Caldwell \etal 1998), which under certain
conditions can correspond to a time-dependent cosmological ``constant''
(e.g. Crooks \etal 2000).

This case is potentially dangerous, as the very equation of state of
dark energy could vary with time. If this occurs and the
variation is significant, a large part of the information
contained in the CMB angular spectrum could have
to be ``spent'' to reconstruct the time dependence of the components
of the stress--energy tensor of dark energy.
Although, in this paper, we do not consider models with time dependent
$\Lambda$, the results obtained with the technique presented here 
 could become
particularly significant for such a case, providing information
complementary to CMB on the time dependence of $\Lambda$.

However, independently of the occurrence of such a
delicate case, it 
should be remarked
that the imprints left 
by density fluctuations on the CMB are memories of a
distant past of the Universe. Data from the local universe can be used to
provide self-consistency tests for structure formation theories tuned at the
matter-radiation decoupling epoch which is probed by CMB observations.

The plan of the paper is as follows.
In the next section we describe the simulations used. In section 3,
we illustrate how we identify the clusters and measure their properties.
In section 4 we work out the statistics of cluster mass quadrupoles and
present our results. Finally, we summarize and discuss our conclusions
in section 5.

\section{The simulations}

We carried out two simulations using the parallel N--body
code of Gardini \etal (1999), which was developed from the serial public
AP3M code of Couchman (1991) extended to different cosmological models 
and different mass
particle sets. 
%
The first simulation was a ``tilted'' 
Einstein-de Sitter model (hereafter TCDM), 
while the second simulation was a $\Lambda$CDM model, i.e. a flat
CDM universe with non--zero cosmological constant.
The parameters of the model are reported in the Table.
Both these models roughly yield the correct abundance of rich
clusters (see \cite{Gar}). The TCDM simulation is the same
already considered by \cite{Gar}; the normalization of the run has been
rescaled to yield $\sigma_8=0.55$ at the final epoch, instead of 
$\sigma_8=0.61$, so that the abundances of rich clusters
are the same for both models at the final epoch.
 
The simulated volumes are
360$\, h^{-1}$Mpc cubic boxes.
CDM+baryons are represented by $180^3$ particles,
whose individual mass is $2.22 \cdot 10^{12} h^{-1} M_\odot$ for TCDM
and $0.777 \cdot 10^{12} h^{-1} M_\odot$ for \lcdm.
We use a 256$^3$ grid to compute the FFTs needed to evaluate the long range
contribution to the force (PM) and we allow for mesh refinement
where the particle density attains or exceeds $\sim 30$
times the mean value.
The starting redshifts are $z_{in}=10$ for TCDM and $z=20$ for \lcdm.
The particle sampling of the density field is obtained applying
the Zel'dovich approximation (\cite{Zel,Dor})
starting from a regular grid.
We adopt the same random phases in both simulated models.

The comoving force resolution is given by
the softening length, $\eta \simeq 112\, h^{-1}$kpc. The force $F$ is evaluated
considering each particle as a smoothed distribution of mass, with shape
$ \rho(r) = (48/\pi \eta^4)(\eta/2-r)$ for $r<\eta/2$ (this is the so--called 
S2 shape, \cite{HE}). It behaves as $F \propto 1/r^2$
when $r \ge \eta$. Since the softening of the force is usually referred
to a Plummer shape, $F \propto r/(r^2+\epsilon^2)^{3/2}$, we used a least 
$\chi^2$ test to estabilish the best approximation between the forces
generated by the two different shapes. The minimum $\chi^2$
occurs when $\eta=2.768\epsilon$. In our case, this corresponds to a Plummer
equivalent softening $\epsilon_{pl} \simeq 40.6 h^{-1}$kpc; we will use this
latter value of the softening as our nominal force resolution.
Our comoving force and mass
resolutions approach the limits of the computational resources of
the machine we used (the HP Exemplar SPP2000 X Class processor
of the CILEA consortium at Segrate--Milan).

The number of steps were
1000 equal $p$--time steps (the time parameter is
$p \propto a^{2/3}$, where $a$ is the expansion factor).
Such step choices were dictated by two criteria: (i) Energy conservation.
According to Layzer Irvine equations (see, e.g., \cite{EDFW}), it
had an overall violation $< 3\, \%$ for TCDM and \lcdm. 
(ii) Cole et al. (1997) requirements, that the rms displacement of particles 
in a step is less than $\eta/4$ and the fastest particle has a displacement 
smaller than $\eta$
were never violated. 

\section{Cluster identification}

One of the main aims of the simulations is that of obtaining
a large set of model clusters for each cosmological model at different
redshifts. This will enable us to study cluster evolution and
to create mock cluster catalogs. 
Here we shall report some basic results and general 
properties of the clusters selected in the simulation outputs.

The clusters we consider were found using a spherical
overdensity (SO) algorithm, yielding the cluster locations, the
radii $R_s$ inside which a density contrast $\delta_{cr} = \delta_{vir}$ is 
attained and the total mass $M$ of the particles within $R_s$. 
As ``virial'' overdensities $\Delta_{vir}$ 
for TCDM and \lcdm we choose the ``standard'' 
values 178 and 110 suggested by the spherical infall model; we
use the relation $\Delta_{vir}=178\Omega_m^{0.45}$ for $\Omega_m+
\Omega_\Lambda=1$ (e.g.~\cite{Eke2}). 
The SO procedure has the benefit of
providing clusters which appear to satisfy a sensible virialization
requirement (a given overdensity in a sphere).

Let us now describe our SO procedure implementation. As a first step,
candidate clusters are located using a standard FoF algorithm, with
linking length $\lambda = \phi \times d$ (here $d$ is the average 
particle--particle separation), yielding groups with more than $N_f$
particles. We then perform the following operations: (i) we find the
center--of--mass $C_M$ of each group and (ii)
we determine the radius $R_g$, inside which the density contrast is 
$\delta_{cr}$ (all particles are included, not only those
initially found by FoF). In general, the new center--of--mass is not $C_M$.
The operations (i) and (ii), define a new particle group, on which
the same operations (i) and (ii) can be repeated. The procedure is
iterated until we converge onto a stable particle set. If, at some stage, the 
group contains less than $N_f$ particles, we discard it. The final $R_g$
is $R_s$. 
It may happen that
a particle is a potential member of two groups; in this case
the procedure assigns it to the more massive one. 
This has the consequence that, sometimes, more massive groups
swallow smaller ones, producing a slight decrease
of the total number of clusters, over all mass scales (see Gardini \etal 1999).
Gardini \etal 1999 describe the SO algorithm and its outputs in detail and 
compare it with those found using
 other group identification algorithms and previous work (e.g. \cite{Gov}).
In this work SO was started setting $\phi = 0.2$ and $N_f$
corresponding to a mass threshold $3.0\times 10^{13}\Msunh$.
Above this mass threshold there are $\sim 10000$ clusters in the samples.
For the analyses, we only use clusters more massive than 
$4.2\times10^{14}\Msunh$ (see \S~5); there are $\sim 300$ of them.

\section{Cluster quadrupoles}

For each cluster identified as above, 
we compute the quadrupole of the cluster particle distribution projected along
a randomly oriented direction through the simulation box. 

If $x_i$ and $y_i$ are the coordinates of the $i$--th cluster particle
when projected onto the plane perpendicular to the axis chosen
(with the origin at the cluster's center), the quadrupole of the projected
particle distribution is defined as:

\be
Q = { (q/N_{vir})^{1/2} \over R_{vir}^2 }
\ee

where 

\be
q = \sum_i^{N_{vir}} \left[ (x_i^2+y_i^2)^2 + 4(x_i y_i)^2 \right]
\ee
The sum extends over the $N_{vir}$ particles contained within $R_{vir}$.

For the analyses presented in the next section, we have repeated
the calculations using three random directions to compute the
quadrupoles and verified that our conclusions do not depend on the projection 
chosen. The results are presented for one projection axis. 

The scales resolved by our simulations are significantly 
larger than those resolved in recent high resolution works
(Brainerd \etal 1997, Moore \etal 1998, 
Ghigna \etal 1998, Klypin \etal 1999, Okamoto \& Habe 1999). 
Therefore, it would be hard to trace in our clusters the
rich mass substructure found by these authors.
It is however unlikely that this affects the
evaluation of a global property such as the quadrupole.
A rough estimate of the resolution needed to obtain accurate
quadrupole measures can be performed by
comparing the linear scale resolved ($r_{res}$) with 
$\pi R/2$ ($R$ is the distance from the center of the
cluster considered). Only the contribution to the
quadrupoles coming from substructure in a central region of radius
not exceeding a few hundred kpc's 
should therefore be affected by lack of resolution.
Let us consider a cluster of radius $R$ and the
contribution to its quadrupole coming from a
central region of projected radius $r$. The ratio between
the volumes of the central region and the whole
cluster is $\simeq (3/2)(r/R)^2$. With $R \sim 1.5\, h^{-1}$Mpc
and $r \sim 150 \, h^{-1}$kpc, such ratio is $\sim 1$--2$\, \%$.
The mass profile is not flat and this increases the weight of the
central region;
however, this bias should not be important because 
the structure of the central regions should be
mostly shaped by the internal dynamics of the cluster,
 with little memory of late tidal actions and/or dishomogeneities
due to secondary infall, which most likely cause
the differences in the global anisotropy of the clusters
in the different models. 

Besides possible numerical effects, there is another limitation
in our analysis. 
The noise that would affect the estimates of $Q$ from the mass 
distribution of real clusters has not been taken into account 
(e.g. from projection effects and from  the limitations of the
techniques used to reconstruct the mass from the observed 
gravitationally induced shear; see Wilson, Cole \& Frenk 1996a).
We plan to improve the analysis along these lines in future work.
However note that the analysis of Reblinsky \& Bartelmann (1999), 
using large scale cosmological simulations,
indicates that projection effects in weak-lensing-mass selected cluster samples
should be small.

\section{Statistical analyses}

For both cosmological models, we have examined the distribution of the
 quadrupoles. As pointed out in the Introduction, these estimators of
cluster morphologies are expected to probe the different formation
histories of clusters in different cosmologies. 
In both cases we select the clusters more massive than 
 $4.2\cdot 10^{14}\Msun$; the numbers of clusters above threshold are 
the same for the two models.



The histogram in Figure~\ref{f:dist_q} shows the overall 
distribution of the quadrupoles measured.
TCDM clusters
have clearly larger values of $Q$ than their \lcdm counterparts, in 
agreement with the expectation of ongoing infall and accretion of
present clusters in the TCDM model. This effect is more prominent 
for very massive clusters, as it is shown in Figure~\ref{f:barre}.  
The figure shows the trend of $Q \,vs.\, M_{vir}$.
Each point is obtained ranging the clusters
in order of increasing mass and averaging over 100 neighbours 
in the lists;
the errorbars correspond to one standard deviation
in the distribution of quadrupoles.

\begin{inlinefigure}
\centerline{\includegraphics[width=1.0\linewidth]{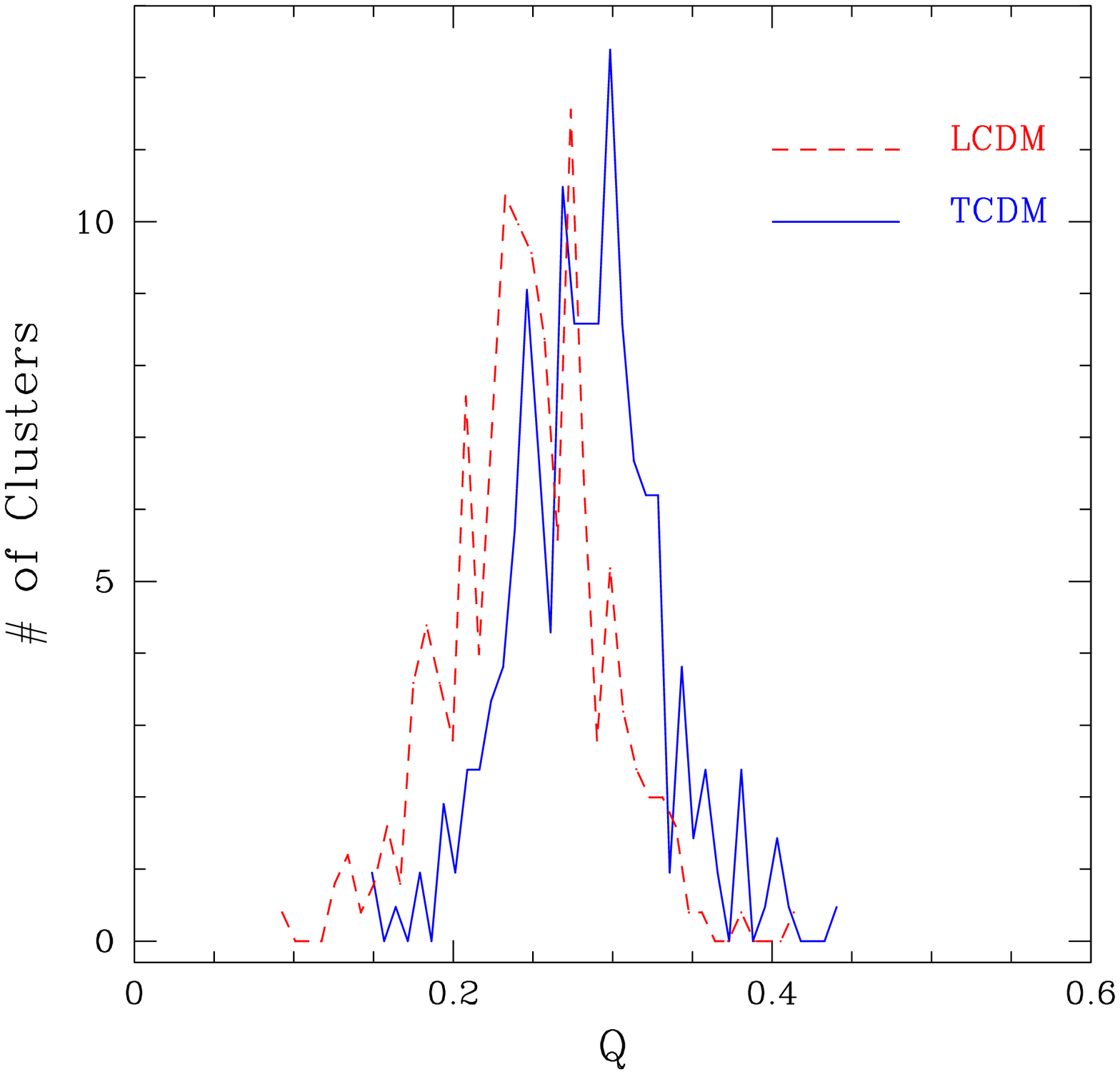}}
\caption{Histogram of the distribution of the quadrupoles
for the whole set of clusters more massive than $4.2\times10^{14}\Msun$
(On the vertical axis we simply plot the number of clusters above threshold 
in each bin). 
TCDM clusters have systematically 
larger quadrupoles than their \lcdm counterparts.
}
\label{f:dist_q}
\end{inlinefigure}


\begin{inlinefigure}
\centerline{\includegraphics[width=1.0\linewidth]{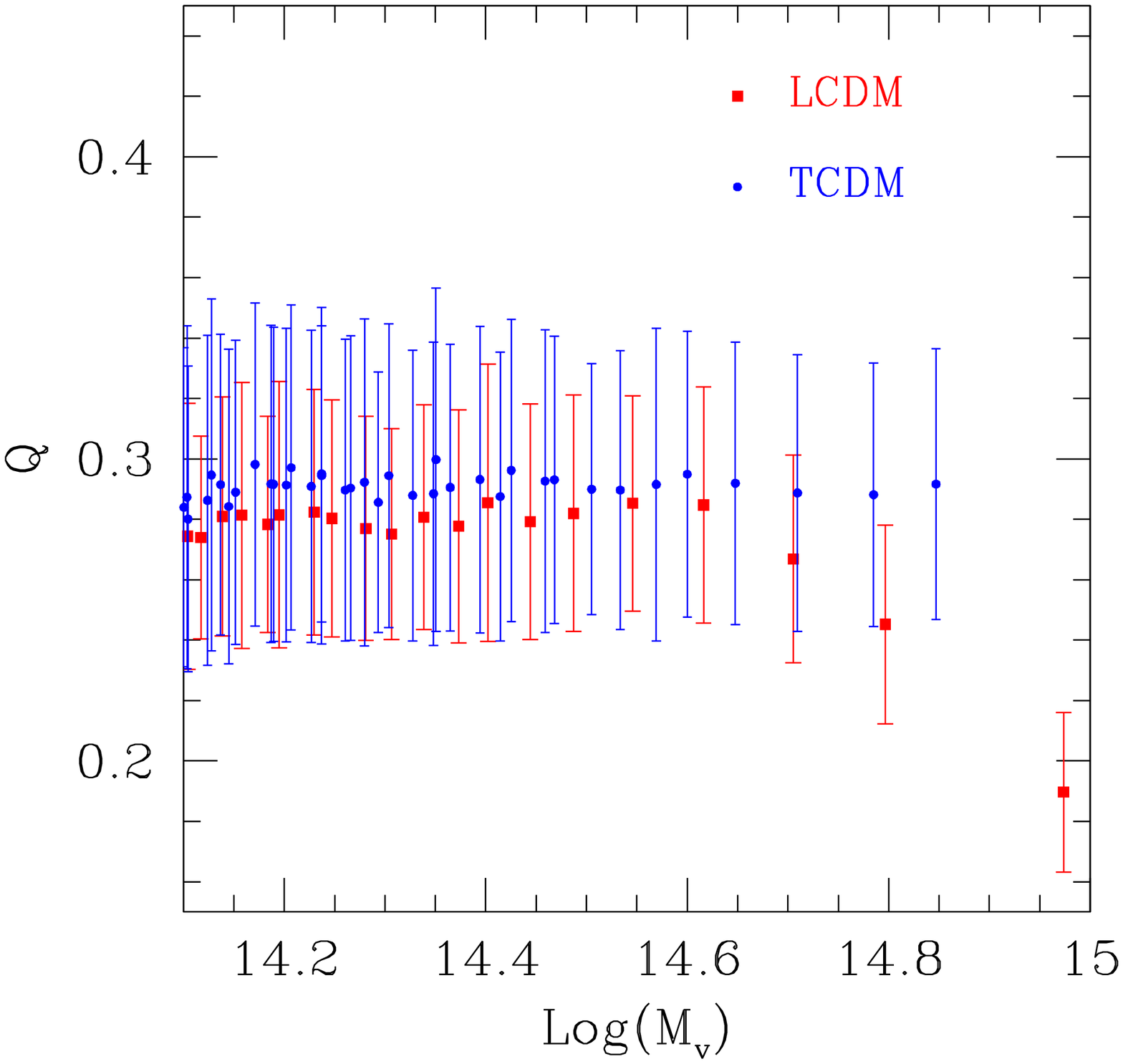}}
\caption{Average $Q$ vs. virial mass $M_{vir}$. The value of $Q$ is
obtained ranging the clusters in order of increasing masses and 
averaging over 100 neighbours (i.e. they are binned 
into bins containing 100 clusters each):
the value $M_{vir}$ reported is the average of the 100 values.}
\label{f:barre}
\end{inlinefigure}

\begin{inlinefigure}
\centerline{\includegraphics[width=1.0\linewidth]{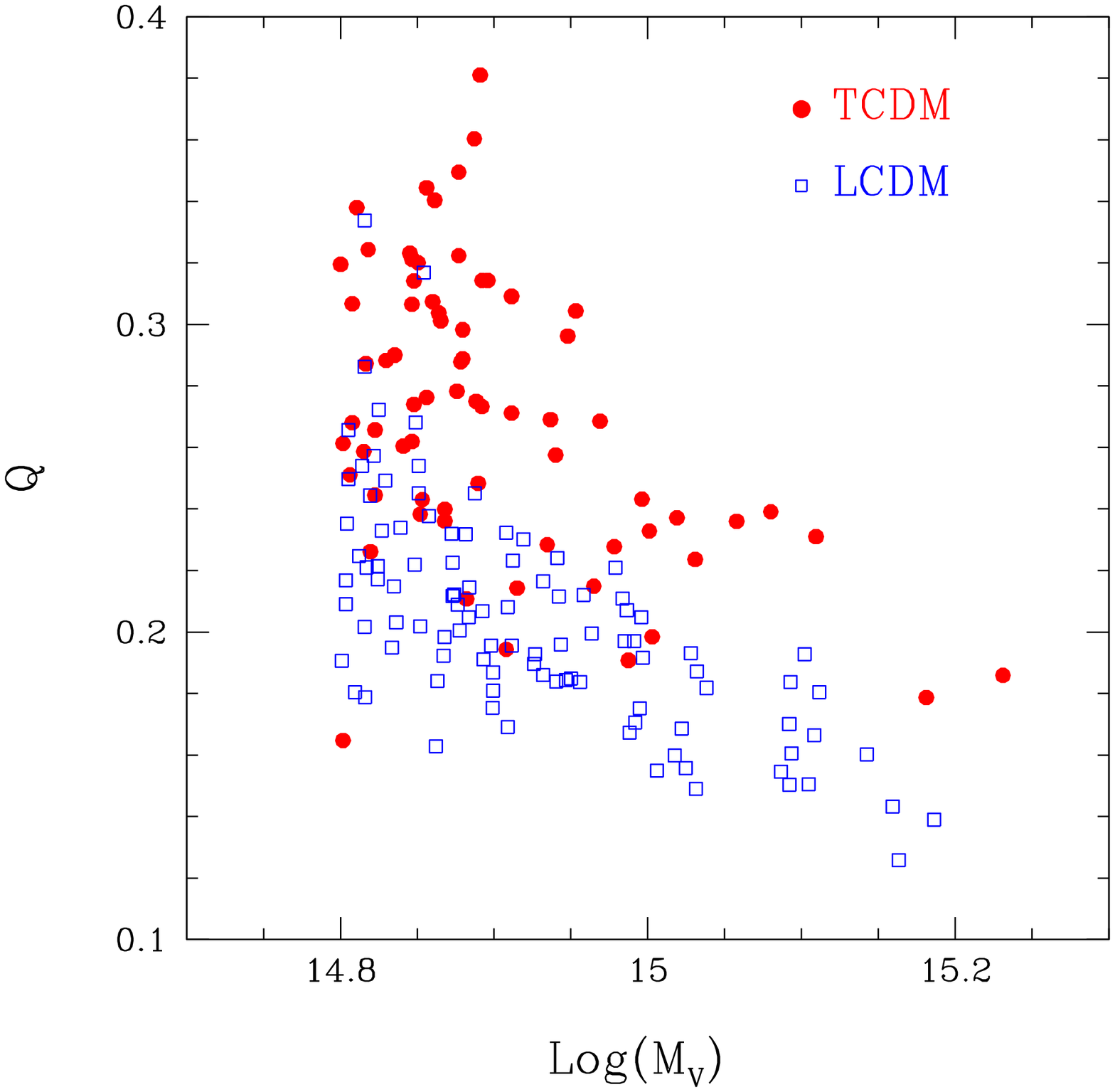}}
\caption{Scatter plot of $Q$ vs. $M_{vir}$ for rich clusters.}
\label{f:quad50}
\end{inlinefigure}

The average quadrupoles of TCDM and \lcdm clusters are 
sistematically different at all masses;
the most interesting result however is the strong difference found at the high
mass end. Note that this effect could not be detected by WCF96, because 
of the small number of simulated clusters in their analysis and, most 
importantly, because they
built a sample of essentially equal mass clusters (they
rescaled the particle masses of their runs so as to obtain same mass objects).
Figure~\ref{f:quad50} is a scatter plot of $Q$ vs $M_{vir}$
for clusters more massive than $6.3\cdot 10^{14}\Msun$, the mass threshold 
for which the average quadrupoles of TCDM and \lcdm clusters start differing 
by several percentual points
 (this threshold yields 105 clusters for \lcdm and 81 for TCDM). 

How significant is the difference between the two distributions ? 
 Could it be detected through observations of real clusters ?
Observations of clusters have made impressive progress in the past decade.
As we have reviewed in the Introduction, maps of the mass distribution itself 
reconstructed from observations of gravitational lensing effects
can be systematically built. On the observational 
side, much progress is a consequence of the development of large--format 
CCD cameras
which can cover fields of $\sim 0.5$ degrees on a side on 4--meter class
telescopes. On the theoretical side, different inversion algorithms
have been developed to systematically exploit the data on weak gravitational 
shear, accounting for the effects of realistic PSFs and optimally weighing
background galaxies (see the recent 
reviews by Mellier, 1999, and Bartelmann \& Schneider, 2000).
Compilations of existing data would allow
samples of $\sim 40$ clusters to be examined (e.g. Mellier 2000). 
However, such 
samples are hardly homogeneous and the analysis of them would require a 
careful examination of systematic effects.
Only very recently an effort to perform a weak lensing study of a large well 
defined sample 
has been started using the NOT and the University of Hawaii 2.24-m
telescope (Dahle 2000).
 On the contrary, inspecting a 
few very massive clusters is a relatively easy task. This is particularly 
welcome for the purposes of the present analysis, given that the most 
prominent
differences between the models appear at the high mass end. 
Let us focus on the behaviour of, say, the first 15 most massive 
clusters in both samples. 
As seen in Figure~\ref{f:quad50} 
TCDM clusters are clearly shifted towards high
values of $Q$.  

\begin{inlinefigure}
\centerline{\includegraphics[width=1.0\linewidth]{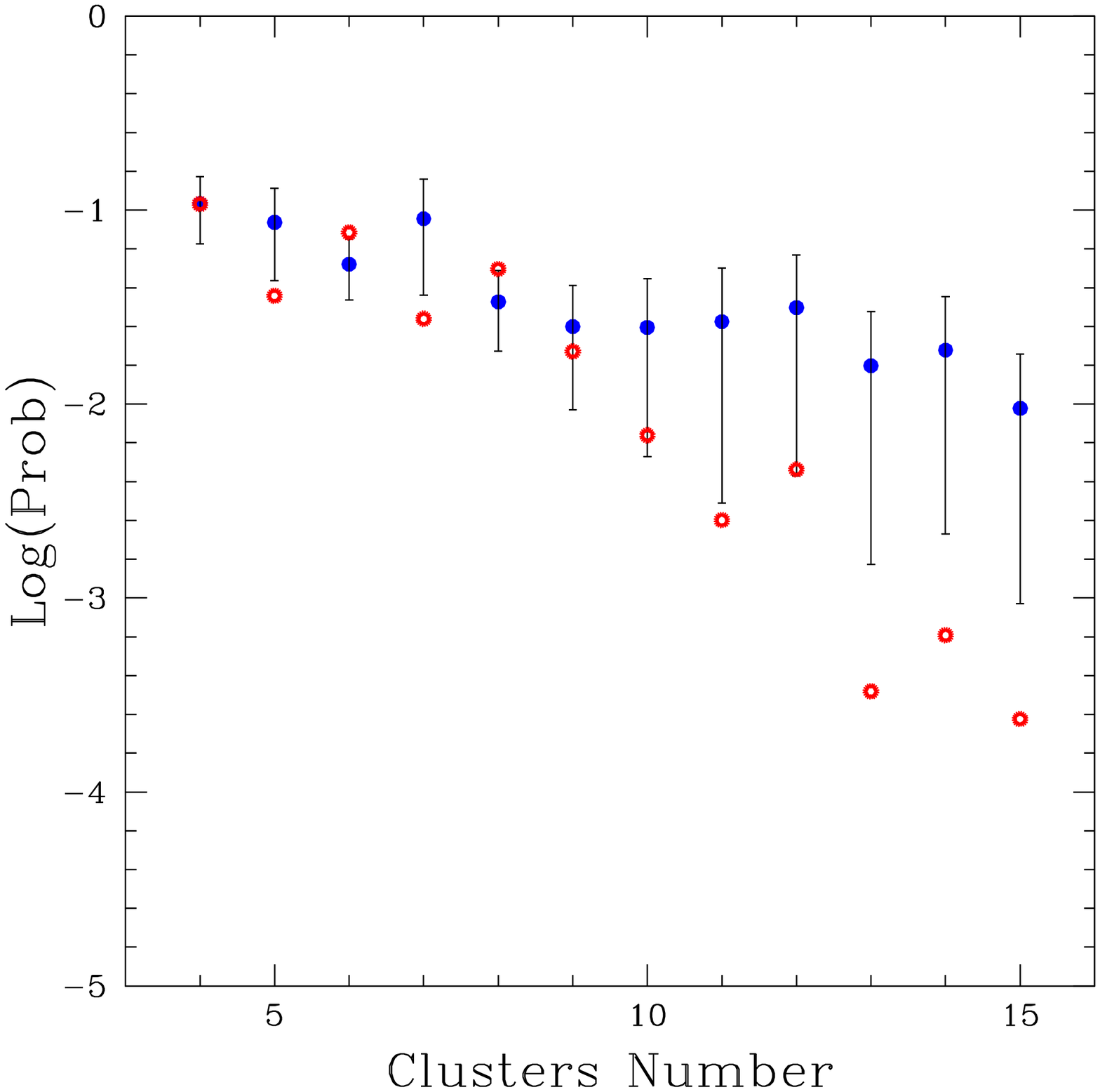}}
\caption{Kolmogorov-Smirnov probability
that the $Q$ measured for the $N$ most massive clusters in the 
simulation volumes are drawn from the 
same theoretical distribution. The (red) dots without errorbars refer to
clusters selected in the whole simulation boxes ($360\Mpch$ a side). 
The (blue) dots
with errorbars are averages over the results obtained for (eight)
half--size subboxes ($180\Mpch$ a side); the errobars correspond to one 
standard deviation.}
\label{f:KS}
\end{inlinefigure}

In order to quantify the significance of this difference, we perform a K-S test
on the distribution of the quadrupoles. Figure~\ref{f:KS} gives the probability
that the $Q$ measured for the $N$ most massive clusters in \lcdm and TCDM 
are drawn from the same underlying distribution (red dots, without errorbars).
 The cluster number $N$ is varied between 4 and 15. Even with only 10 clusters
the probability that the difference between the $Q$ measured for the \lcdm 
clusters and their TCDM counterparts is just a statistical fluke is below 1\%; 
using 15 clusters, it is well below $1/1000$.
Extending the analysis to all the clusters more massive than $ 6.3\cdot 
10^{14}\Msun$ (about twice the mass of the Virgo cluster)
 yields a KS probability of a mere $10^{-32}$.

The simulated clusters live in a cubic volume of $360\Mpch$ a side. 
According to the result above, picking 
$\sim 10$ massive clusters ($M_{vir}\sim 10^{15}\Msun$) 
in such a volume is
sufficient to discriminate between the two models at the 1\% confidence
limit. Of course, the result will depend on the volume explored, both because 
$10^{15}\Msun$ clusters are rare objects and because of cosmic variance.
Ideally, 
different simulations would be welcome to thouroughly examine this issue.
 As a tentative test, 
we have divided the simulation boxes in eight subboxes of $180\Mpch$ a side;
for each pair of corresponding \lcdm and TCDM subboxes, we have then repeated 
the same analyses as for the parent
boxes, i.e. we have selected 
the 15 most massive clusters and estimated the KS 
probabilities that their $Q$ distributions are drawn from the same underlying
model. The results are shown in figure~\ref{f:KS} as blue dots with errorbars
(1--$\sigma$).
The average KS probability increases markedly; however the result is
dominated by the outcome of the KS analysis in one of the subboxes, in which
the most massive clusters have very similar distributions, whereas in the
other subboxes the KS test yields probabilities only slighlty higher than
for the whole box. 
This indicates that at least a volume $\gsim 200 \Mpch$ is necessary
to obtain robust quadrupole statistics able to discriminate between the
two cosmologies just using a few tens of massive clusters. 
A set of large volume simulations
would be welcome to examine this issue further. 

\section{Discussion and Conclusions}

Let us first summarize the analyses reported in this paper. 
We have used two large volume simulations of a \lcdm and a TCDM universe, which
were started from the same initial random numbers so that the structure
forming in each of them are directly comparable. The simulations span a cubic 
volume of $360\Mpc$ a side. The numerical parameters of the simulations and
the linear features of the cosmological models are examined in \S~2. 

In each simulated volume we have identified the massive virialized systems that
would correspond (in mass) 
to observed galaxy clusters. We have then measured the 
quadrupole $Q$ of the projected mass distribution of each simulated cluster
(we extend our analysis to objects more massive than a threshold 
$=\,4\times 10^{14}\Msun$).
 
Studying the statistics of the quadrupoles, we have obtained the following
results:

\begin{itemize}

\item{}
The distributions of the quadrupoles of TCDM and \lcdm clusters differ 
significantly, with TCDM clusters having systematically 
larger quadrupoles in agreement 
with the linear theory expectation that in a high density universe clusters
keep accreting material till the present epoch while in a low density universe
late accretion and mergers are suppressed. 
This confirms previous numerical results 
by Evrard \etal (1993)
Mohr \etal (1995) and Wilson, Cole \& Frenk (1996b).

\item{} 
At variance with the previous authors which focussed on a small
set of (at that time) high resolution simulations, we have built
catalogs of simulated clusters. This allows us to study the
dependence of the quadrupoles on cluster mass. We find that the trends 
of $Q$ vs. $M_{vir}$ of the two models are markedly different
at the high mass end. For
$M_{vir}\gsim8\cdot 10^{14}\Msun$ (about twice the mass of the Virgo cluster), 
the average quadrupoles of \lcdm clusters drop to values 30\% lower than
those of their TCDM counterparts of similar mass. 

\item{}
We have examined how many massive clusters are needed to significatly 
distinguish between the quadrupole
distributions of TCDM and \lcdm clusters performing a KS test.
We find that, using just the 10 most massive
clusters of the samples ($\gsim 10^{15}\Msun$ in the simulation
 volume of $360\Mpch$ a 
side), 
the probability that the $Q$ measured are
drawn from the same model distribution is less than 1\%, and decreases
to well below $1/1000$ if the 15 most massive clusters are employed.

\item{}
The significance of the test depends on the volume span by the cluster
sample and on the number of {\it most--massive} clusters which
can be taken into account. If such number is restricted to
$\cal O$(10), we estimate that a volume $\gsim 200\Mpch$ is necessary
to obtain robust quadrupole statistics able to discriminate between the
two cosmologies. Of course, increasing the number of clusters (that is 
lowering the mass/richness threshold for a cluster to be considered)
implies more and more 
careful observational work. It should also be noted that, if
the mass threshold is pushed down to
 $\lsim 4\times10^{14}\Msunh$, no substantial advantage
would result from adding more clusters to the sample, as,
at such masses, the signal tends to quickly fade. 
For example,
in a box with side $\sim 180 \, h^{-1}$Mpc, we find
$\sim 25$ clusters above such threshold.
In a volume $\sim 10^7$--$10^8 h^{-3}$Mpc$^3$, of the order of what 
is observationally inspected, the number of {\it significant} clusters
approaches a total of $\sim 200$. A set of new
simulations could be used to determine how effectively different values of
$\Lambda$ can be discriminated, should this whole
sample of potential data be exploited. 
From our analysis, we speculate 
that it could allow us to go beyond
discriminating between $\Lambda = 0$ and $\Lambda \neq 0$.

\end{itemize}

These results confirm that the statistics of cluster quadrupoles
is sensitive to the cosmological model. 
 In particular this statistics
could do more than distinguishing between the 
currently popular \lcdm model and a standard
CDM cosmology with zero cosmological constant. 
Even in the era of high precision cosmology that CMB observations
will be hopefully opening soon, the statistical analysis
presented here might be quite
significant. First of all, it conveys information coming from
our own or nearby epochs, allowing a self consistency test
of theories tuned at the matter--radiation decoupling epoch. 
Furthermore, in the case of complex models of dark energy as discussed in the
Introduction,
it can complement CMB observations, helping to remove residual
degenerations in the determination of the cosmological parameters.
 Of course, using quadrupole statistics requires, 
from the observational point of view,
a strong systematic effort 
to expand the available data set on weak lensing from
galaxy clusters,
and, from the theoretical side, a careful examination of the 
biases affecting the quadrupole estimates of real clusters and their
selection (e.g. due to projection effects or the limitations of cluster
identification and mass reconstruction algorithms; see e.g. Wilson, Cole \& 
Frenk 1996a and Reblinsky \& Bartelmann 1999).   

\acknowledgments
We thank the consortium CILEA for allowing us to run a
former trial simulation free of charge and a particular thank is
due to Giampaolo Bottoni of CILEA, for his expert technical assistance.


\vfill\eject

\begin{table}
\begin{center}
\begin{tabular}{l|c|c}
\multicolumn{3}{c}{} \\
&\multicolumn{1}{c}{TCDM}&\multicolumn{1}{c}{\lcdm}
\\
\hline \\
 $\Omega_m$ &      1   & 0.65\\
 $\Lambda$ &       0   & 0.35\\
 $\Omega_b \cdot 10^2$ &      6   & 6\\
 $n$                 &     0.8  & 1.05\\
 $h$ &       0.5   & 0.65\\ 
 $Q_{PS,rms}/\mu$K      &    15.6  & 20.05 \\
 $\sigma_8$            &    0.55  & 1.08 \\
 $\Gamma$             &   0.32  & 0.19  \\
 $N_{cl}$ (PS; $\delta_c = 1.69$)   &   4.2 & 4.2\\
\multicolumn{3}{c}{} \\
\multicolumn{3}{c}{} \\
\end{tabular}
\caption{
Parameters of the models. All parameters listed are
either input parameters or quantities worked out from the
linear theory (see \cite{Gar}). 
The normalization to COBE quadrupole was deliberately kept at the
$\sim 3 \, \sigma$ lower limit, in order to leave some room to the
contribution of tensor modes, while being consistent with the data.
The expected interval for $N_{cl} $ is 4--6.
}
\label{tab1}
\end{center}
\end{table}

\end{document}